\documentclass[aip, amsmath, amssymb, nobibnotes, nofootinbib, citeautoscript, reprint, superscriptaddress]{revtex4-1}
\usepackage{graphicx}
\usepackage{dcolumn}
\usepackage{bm}
\usepackage[version=3]{mhchem}
\usepackage{multirow}
\usepackage{sourcecodepro, sourcesanspro, sourceserifpro}
\usepackage[T1]{fontenc}
\usepackage[protrusion=true,expansion=true,final]{microtype}
\usepackage{mathastext}
\usepackage[usenames,dvipsnames]{xcolor}
\usepackage[bookmarks=true,colorlinks]{hyperref}
\hypersetup{
	linkcolor= blue,        
	citecolor=blue,        
	filecolor=blue,        
	urlcolor=blue,         
	pdfauthor={Debangshu Mukherjee},
	pdftitle={A roadmap for edge computing enabled automated multidimensional transmission electron microscopy},
}
\usepackage{xfrac}
\usepackage{url}

\makeatletter
\def\blfootnote{\gdef\@thefnmark{}\@footnotetext}
\makeatother

\begin{document}
	\title{A roadmap for edge computing enabled automated multidimensional transmission electron microscopy}
	\author{Debangshu\ Mukherjee}
	\email{mukherjeed@ornl.gov}
	\affiliation{Computational Sciences \& Engineering Division, Oak Ridge National Laboratory, Oak Ridge, Tennessee 37831, USA}
	\author{Kevin M.\ Roccapriore}
	\affiliation{Center for Nanophase Materials Sciences, Oak Ridge National Laboratory, Oak Ridge, Tennessee 37831, USA}
	\author{Anees\ Al-Najjar}
	\affiliation{Computational Sciences \& Engineering Division, Oak Ridge National Laboratory, Oak Ridge, Tennessee 37831, USA}
	\author{Ayana\ Ghosh}
	\affiliation{Computational Sciences \& Engineering Division, Oak Ridge National Laboratory, Oak Ridge, Tennessee 37831, USA}
	\affiliation{Center for Nanophase Materials Sciences, Oak Ridge National Laboratory, Oak Ridge, Tennessee 37831, USA}
	\author{Jacob D.\ Hinkle}
	\affiliation{Computational Sciences \& Engineering Division, Oak Ridge National Laboratory, Oak Ridge, Tennessee 37831, USA}
	\author{Andrew R.\ Lupini}
	\affiliation{Center for Nanophase Materials Sciences, Oak Ridge National Laboratory, Oak Ridge, Tennessee 37831, USA}
	\author{Rama K.\ Vasudevan}
	\affiliation{Center for Nanophase Materials Sciences, Oak Ridge National Laboratory, Oak Ridge, Tennessee 37831, USA}
	\author{Sergei V.\ Kalinin}
	\affiliation{Department of Materials Science \& Engineering, Tickle College of Engineering, University of Tennessee, Knoxville, Tennessee 37996, USA}
	\affiliation{Currently at: Special Projects, Amazon Science}
	\author{Olga S.\ Ovchinnikova}
	\affiliation{Computational Sciences \& Engineering  Division, Oak Ridge National Laboratory, Oak Ridge, Tennessee 37831, USA}
	\affiliation{Currently at: Division of Systems Engineering, ThermoFisher Scientific, Bothell, Washington 98021, USA}
	\author{Maxim A.\ Ziatdinov}
	\affiliation{Computational Sciences \& Engineering  Division, Oak Ridge National Laboratory, Oak Ridge, Tennessee 37831, USA}
	\affiliation{Center for Nanophase Materials Sciences, Oak Ridge National Laboratory, Oak Ridge, Tennessee 37831, USA}
	\author{Nageswara S.\ Rao}
	\affiliation{Computational Sciences \& Engineering Division, Oak Ridge National Laboratory, Oak Ridge, Tennessee 37831, USA}

    \date{\today}
	
	\begin{abstract}
		The advent of modern, high-speed electron detectors has made the collection of multidimensional hyperspectral transmission electron microscopy datasets, such as 4D-STEM, a routine. However, many microscopists find such experiments daunting since such datasets' analysis, collection, long-term storage, and networking remain challenging. Some common issues are the large and unwieldy size of the said datasets, often running into several gigabytes, non-standardized data analysis routines, and a lack of clarity about the computing and network resources needed to utilize the electron microscope fully. However, the existing computing and networking bottlenecks introduce significant penalties in each step of these experiments, and thus, real-time analysis-driven automated experimentation for multidimensional TEM is exceptionally challenging. One solution is integrating microscopy with edge computing, where moderately powerful computational hardware performs the preliminary analysis before handing off the heavier computation to HPC systems. In this perspective, we trace the roots of computation in modern electron microscopy, demonstrate deep learning experiments running on an edge system, and discuss the networking requirements for tying together microscopes, edge computers, and HPC systems. 
	\end{abstract}
    \maketitle
	
	\section{\label{sec:intro}Introduction}
	\blfootnote{\textsf{This manuscript has been authored by UT-Battelle, LLC under Contract No. DE-AC05-00OR22725 with the U.S. Department of Energy. The United States Government retains and the publisher, by accepting the article for publication, acknowledges that the United States Government retains a non-exclusive, paid-up, irrevocable, world-wide license to publish or reproduce the published form of this manuscript, or allow others to do so, for United States Government purposes. The Department of Energy will provide public access to these results of federally sponsored research in accordance with the DOE Public Access Plan (\href{http://energy.gov/downloads/doe-public-access-plan}{http://energy.gov/downloads/doe-public-access-plan})}}
	Ernst Ruska built the first transmission electron microscope (TEM) during his doctoral studies, and it celebrates its eightieth anniversary this year\cite{ruska_1,ruska_2, originTEM}. Interestingly, this system was built and was operational less than a decade after experimental results from Davisson and Germer proved de Broglie's hypothesis to be correct about the dual wave-particle nature of electrons \cite{davisson_germer, deBroglie}. Optical microscopes inspired the first TEM, and since then, several new imaging modalities have been implemented, such as electron holography\cite{gabor_holography, holography_20}, Lorentz electron microscopy\cite{lorentz_em}, scanning TEM (STEM)\cite{hrstem, stem_atoms} to name a few. Even with the introduction of new imaging modalities, the electron microscope developed by Knoll and Ruska is still remarkably similar to machines still in use today. However, in these past eight decades, TEMs have continued to gain capabilities such as energy dispersive X-ray spectroscopy (EDX)\cite{castaing_thesis}, electron energy loss spectroscopy (EELS), annular dark field (ADF) detectors, and aberration-corrected optics both in the TEM and STEM modes.  
	
	These advancements have allowed the S/TEM to play a crucial role in analyzing nanometer-scale structural phenomena in physical and life sciences. It has helped tie together the structure-property relationships in materials systems as diverse as interfaces\cite{rp_stone}, superlattices \cite{polar_vortices}, domain walls\cite{nelson_domains, linbo3_domain}, grain boundaries\cite{grain_boundary}, nanoparticles\cite{atomic_tomo}, catalyst surfaces\cite{hrsem}, etc. This has led to the discovery of novel applications such as phonon modes at polar vortices, two-dimensional electrical liquids at oxide interfaces, etc\cite{polar_vortices_eels, 2deg, stem_review}. In the physical sciences, these applications have been in fields as wide as lithium-ion batteries, catalyst systems, alloy designs to integrated electronic circuits, to name a few\cite{battery_cryoTEM, catalysis_joule, alloy, IC_TEM, IC_TEM_2}. Hardware advancements in electron microscopy, such as specialized holders for cryogenic, heating, biasing, or liquid-cell work over the past few decades, have also enabled nanoscale studies of dynamic systems such as materials under mechanical strain, thermal gradients, switching behavior in oxide ferroelectrics or reaction phase systems such as catalysts. So extensive has been the advancements that microscopists have rightly pointed out that the modern transmission electron microscope with its capabilities for electron imaging, electron diffraction, spectroscopy, operando studies is ``A Synchrotron in a Microscope''\cite{synchrotron_tem,synchrotron_tem20}.

	Along with the massive advancements in physical sciences - the transmission electron microscope has been an absolute game changer for observing biological tissue. Biological TEM followed the development of the original electron microscope closely, with Helmut Ruska, who was Ernst Ruska's brother coincidentally, using the TEM for imaging bacteria and viruses as early as 1939\cite{biological_tem_first}. The development of cryogenic sample processing and microscopy, combined with ultra-fast high sensitivity direct electron detectors, have opened up the entire world of biological systems such as virus cells or individual proteins. As the global coronavirus pandemic continues its' onslaught worldwide, one of the most widely circulated electron microscopy images has been the cryo-EM image of the SARS-CoV-2 virus.
	
	\section{\label{sec:compute}The transition to compute-enabled electron microscopy}
	
	However, as Kirkland observed, advances in microscopy and computers happened almost independently of each other for the first few decades of electron microscopy, with the computation most commonly used for simulating electron microscopy images\cite{kirkland_book}. This situation persists to a certain extent, and a significant portion of electron microscopy, as practiced even today, is thus often anecdotal and susceptible to operator bias. Electron transparent samples require significant manual input for their preparation, and their regions of interest to be imaged are still being decided by the microscopist. This region is often chosen based on the visual intuition of a trained microscope operator and then the data collected from the said region. While automated sample preparation and interfacing with electron microscopes are outside the purview of the current perspective, this current state of affairs is a direct consequence of the fact that microscopy and its associated analysis have continuously operated in the storage and computation-constrained world. Thus, it fell upon a trained microscopist to choose which areas to image and which images to analyze to be judicious with their limitations.
	
	Since then, several changes have made microscopy and computation much more closely coupled with each other over the past two decades. First, several individual lens parameters underpinning values such as aberration or astigmatism have been abstracted away due to the computer-controlled operation of individual microscope components. This abstraction was necessary due to the advent of aberration-corrected electron optics, which iteratively minimized lens aberrations through a combination of multiple quadrupole, hexapole, and octapole lenses; thus, correcting individual lenses becomes tiresome and error-prone. Second, electron microscopy has almost entirely moved away from using photographic film as the data acquisition media towards electronic CCD and CMOS-based detectors.
	
	As a result, S/TEM alignment, operation, and data collection have become significantly automated in recent years. The volume of data that can thus be generated in a single day of microscopy can often be several terabytes. Current generation fast direct electron detectors can generate tens of gigabytes of data in a minute. Similarly, in situ experiments with modern detectors, which often have 4k pixels along a dimension, are often run for several hours and generate hundreds of gigabytes of data per hour. Currently, very few microscopy facilities have on-site computational capabilities to compress, process, and archive such data streams from the microscopes in real-time, let alone use that data for decision-making to drive automated experiments. Several recent publications, notably a recent perspective by Spurgeon et al.\cite{nmat_dataEM_perspective} have raised the issue of the volume of data flooding out of modern electron microscopes and the communities' scattered responses in dealing with the issue.
	
	Thus, there is a need for integrating on-site microscope facilities with computing and storage systems, at the local or remote edge, to form seamless ecosystems, wherein significant measurements collection and instrument steering operations can be automated and remotely orchestrated. In the coming sections, we will give a brief overview of the data deluge in electron microscopy, discuss in brief current computational efforts in the field and elucidate the path forward for edge computing infrastructure for electron microscopy in the materials community.

	\subsection{\label{ssec:detectors}Multidimensional Electron Microscopy Enabled from Detector Advances}
	
	Electronic detectors have been used for TEM image acquisition since the early nineties. For a long time, such detectors were charge-coupled devices (CCD). However, these detectors were indirect, as the CCDs would not record the electrons themselves. Instead, the electrons would interact with scintillators. The scintillators would convert electrons to photons, which would be transferred to the detector through fiber optics that coupled the detectors with the scintillators. However, such a setup degrades the detector quantum efficiency (DQE), and blurs the detector point spread function (PSF) for electron detection. Issues with scintillators are present for X-Ray photon detection too, and thus over the last two decades, as a replacement - semiconductor-based direct detectors have been designed, initially for synchrotrons to detect X-ray photons, and then subsequently for electron detection. Direct electron detectors record individual electron impingement events without any intermediate conversion to photons through scintillators and thus mitigate the DQE and PSF issue with scintillator coupled detectors. A side effect of direct electron detectors is that, along with DQE, the point spread function (PSF) also improves. This impetus for direct electron detectors was from the biological cryo-EM community, where the samples are often highly susceptible to electron dose rates, with the maximum allowable dose often below 10 $e^{-}$/\AA$\mathrm{^2}$.
	
	Along with the improvement in detection capabilities, another focus area of research has been faster detectors. Again, this was partly driven by cryo-EM, as samples degrade rapidly when exposed to electrons, and thus speed is necessary. Modern direct electron detectors employed for 4D-STEM experiments can currently capture over 10,000 frames per second, with the camera developed at Berkeley Lab capable of 87,000 frames per second\cite{ncem_camera}. 

	After speed and sensitivity, the third focus area of electron detector research is ``dynamic range''. Dynamic range refers to the ratio between the highest and lowest electron flux that can be detected simultaneously. Often, detectors that optimize for detection at low electron counts all the way till detection of individual electron impingement events have a lower absolute dose limit. While a detailed discussion about detector geometries is out of the scope for this perspective, dynamic range issues can be solved to a large extent by using hybrid-pixel array detectors\cite{dectris}. The first such detector used for electron microscopy was the Medipix detector, which was spun out from the detector work at the Diamond Light Source Synchrotron facility in the United Kingdom\cite{medipix}. The second such effort, also originally an outcome of synchrotron detector work, is the Electron Microscope Pixel Array Detector (EMPAD), developed at Cornell University\cite{empad1, empad2}. The EMPAD family of detectors was specifically optimized for high dynamic range (HDR), with EMPAD2 reaching a 100,000:1 range for detection. HDR detectors have advantages in both EELS and 4D-STEM experiments; since, in both cases, the ratio between scattered and unscattered electrons may be very high. 
	
	4D-STEM datasets obtained from EMPAD detectors have twice broken the resolution limit in electron microscopes, at 0.4\AA\ in 2018\cite{ptycho_deepsuba}, and 0.25\AA\ in 2021\cite{ptycho_lattice} - through a technique known as electron ptychography where the elastically scattered electron diffraction patterns (the 4D-STEM dataset) is used to solve for the microscope lens parameters and the transfer function of the sample being imaged. The second result reached the physically possible resolution limit before thermal vibrations from atom columns overtake lens aberrations as the primary source of blurring\cite{ptycho_lattice}. These ptychography results have demonstrated that given enough redundancy in the collected 4D-STEM data, it is possible to completely deconvolve the electron lens transfer functions, probe decoherence, and positioning errors from the dataset to generate the pure material transfer function. As a result, the final image quality is significantly better than what can be obtained through classical aberration-corrected ADF-STEM imaging, with the added advantage of requiring lower electron dose rates\cite{mixed_ptycho}.
	
	Because of these advantages, the last two decades have seen electron microscopes retrofitted worldwide with faster, direct electron detectors, not only for imaging but also for EELS and 4D-STEM experiments. These advancements have made the modern STEM truly multidimensional and multimodal, combining imaging, diffraction and spectroscopy in a single equipment. Since then, some of the most significant advancements in electron microscopy have been possible due to the advent of high-speed direct electron detectors with DQE values approaching unity and combined with single electron detection sensitivity \cite{faruqi_review, faruqi_direct}.
	
	\subsection{\label{ssec:data_analysis}Quantitative analysis from electron microscopy datasets}
	
	The advancements in microscopy hardware have made the extraction of quantitative material information from electron micrographs possible. Several recent open-source software packages have been developed by microscopists worldwide to enable this. Some examples include STEMTool\cite{stemtool, stemtool_package}, py4DSTEM\cite{py4dstem}, Pycroscopy\cite{pycroscopy}, PyXem\cite{pyxem}, pixSTEM\cite{fast_detectors1, fast_detectors2}, LiberTEM\cite{libertem} to name a few. Each of these packages focuses on a specific area of TEM analysis, such as Pycroscopy's focus on image processing or py4DSTEM's focus on 4D-STEM data analysis. The most common area of software development appears to be 4D-STEM, with pixSTEM, LiberTEM, and even STEMTool focusing primarily on it. This focus on 4D-STEM is probably driven by the fact that such datasets are often not human parsable, and thus computational analysis is essential to make sense of such datasets. However, since the modern STEM is effectively a highly multimodal equipment, many other large multi-gigabyte datasets are also routinely generated, such as spectral maps from EELS or EDX, long-duration in-situ TEM experiments etc. 
	
	In single particle cryo-STEM, too, large datasets (several hundred gigabytes uncompressed size) are routinely captured before alignment and particle picking. A brief perusal of the Electron Microscopy Public Image Archive (EMPIAR)\cite{empiar} will turn up hundreds of such datasets, each of which individually can be several terabytes. The cryo-EM community, however, has converged on a few open-source solutions such as Relion\cite{relion} or commercial software such as cryoSPARC\cite{cryosparc} for particle reconstruction from images, while the materials science community is more diverse in its' software choices. 
	
	Along with software development, advances in computational capabilities, including accessible CPU/GPUs, implementations of algorithms, and physical models, have led to significant developments in the fields of computational simulations spanning over a range of time and length scales. Therefore, using either experimental or simulated data or both to construct Artificial Intelligence (AI) and Machine Learning (ML) based frameworks for analyzing microscopy datasets have become common in recent years.
	
	While many such studies involve utilization of already developed classification or regression algorithms, frameworks to appropriately find features of interest (such as atoms, defects) from microscopic images, capturing dynamic behavior of the systems, and finally porting them to simulations environment for a comprehensive understanding of the material, are still in its infancy. Even though the primary software used for analysis codes, such as PyTorch\cite{pytorch}, TensorFlow\cite{tensorflow}, Scikit-Learn\cite{scikit-learn}, and JAX\cite{jax}, are all open-access, materials- or problem-specific workflows are often not made public. However, the learning curve required for adapting to one of such existing workflows can be challenging and time-consuming. This is applicable even for simulation packages, most of which are developed in C++, FORTRAN, and Java programming languages, that may not be straightforward to adapt. Alternative ways of using these pieces of software rely on open-source post-processing codes. A non-exhaustive list of available and widely used tools includes p4vasp, vasptools, OVITO\cite{ovito}, Atomsk\cite{atomsk}, Packmol\cite{packmol}, Avogadro\cite{avogadro}, ASE\cite{ASE}, LIGGGHTS, ParaView, PyMol, VMD, Vesta, and many more. The challenge of making these flexible and extensible to process various outputs generated in different formats with varying accuracies, along with integrating them with an experimental setup remains to be addressed.
	
	\section{\label{sec:edge_em}Towards edge-computing in electron microscopy}
	
	Because of the multiple modalities of the data being generated from the electron microscope, there is often huge diversity in the data generated in terms of size, shape, generation rates etc. Thus, there is no one single computational solution for analyzing and automating microscopy experiments. Microscope simulations, and training can occur on high-performance computing (HPC) systems, while initial analysis and trained models should be run on connected edge computational systems, which have low latency between the microscope detectors to the computer memory. In computer hardware, HPC refers to systems that are composed of multiple, often thousands of CPUs and GPUs. These are often monolithic systems, with megawatts of power consumption. HPC systems are very rarely used all at once by a single program, and are often used by multiple groups running multiple programs simultaneously. Compared to HPC, edge computing is a more recent term, and refers to systems intermediate in capability between a HPC system and a desktop. These systems often have multiple GPUs, with a power draw of several kilowatts. The system being currently deployed at ORNL, have Nvidia DGX boxes as the edge systems, while our HPC system is Summit, currently the fourth fastest computer globally with a peak of 200 petaflops. We plan on transitioning to Frontier in the coming few months, which is currently the fastest system globally and the only exascale computer in existence.
	
	\subsection{\label{ssec:automated_em}Ongoing edge computing enabled automated electron microscopy experiments in the physical sciences}
	\begin{figure*}
		\includegraphics[width=\textwidth]{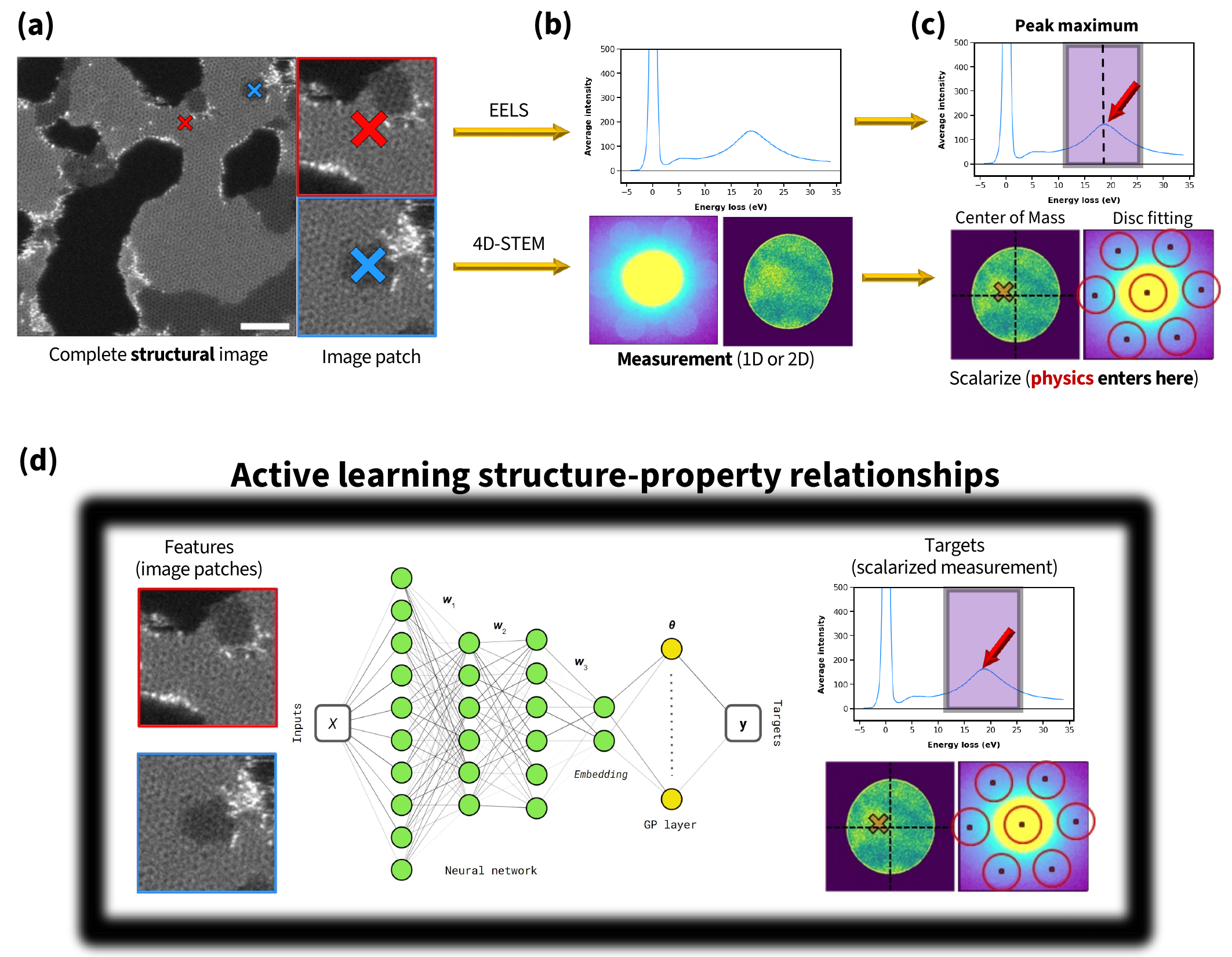}
		\caption{\label{fig:dkl}\textbf{Deep Kernel Learning from microscopy data. (a)} Obtained atomic-resolution image through HAADF-STEM in the electron microscope, which is used as the starting point for patch identification. Multiple varieties of image patches exist, from pure \ce{MoS2}, to \ce{MoS2} with dopants. \textbf{(b)} Multimodal datasets accompany the atomic resolution HAADF-STEM image, such as simultaneously collected EELS or 4D-STEM datasets. \textbf{(c)} The multimodal datasets are scalarized to develop relationships with the imaging data. This scalarization can be of the peak mazima for EELS spectra or center of mass for 4D-STEM diffractograms. \textbf{(d)} Thus, an active learning module can be built to reside on a connected edge computing system, that takes in the features from the HAADF-STEM images and relates them to the scalar feautures in the correlated multimodal datasets (EELS or 4D-STEM) in this case. Adapted with changes from \href{https://pubs.acs.org/doi/10.1021/acsnano.1c11118}{Roccapriore et al. ACS Nano (2022).}}
	\end{figure*}
    A typical experimental workflow in STEM proceeds as follows. First, a 2-dimensional (HA)ADF image is obtained over a relatively large area. Then, a human operator selects regions of interest for more in-depth exploration based on the prior knowledge and specific phenomena they are interested in (e.g., properties around dopants and defects) or, in some cases, a simple ``educated guess''. Here, the ``in-depth'' exploration refers to acquiring a 1D or 2D spectroscopic signal in designated pixels of the 2D image that measures a functionality (`property') of interest, as demonstrated in \autoref{fig:dkl}(a)-(c). Alternatively, one can measure a spectroscopic signal in every pixel of the original image. However, the behaviors of interest are typically localized in relatively small regions of the sample. The entire grid scan is time-consuming, usually imparts a very high electron dose to the specimen, and is only reasonable without prior knowledge about the system. 
    
    Such a setup can be used for automated analysis of multidimensional datasets. Ophus et al. have demonstrated this year, that the py4DSTEM package can be integrated with microscope data acquisition for automated crystallographic orientation mapping\cite{automated_py4dstem}. In another example, it was demonstrated that the obtained 4D-STEM datasets can be used for performing single side-band (SSB) ptychography\cite{ssb_ptycho}, in real time by using GPU based solvers on a connected edge system\cite{real_time_ptycho_ieee, real_time_ptycho_mm}.
    
    Recently at ORNL, we demonstrated a probabilistic machine learning workflow for the intelligent sampling of measurement points toward physics discovery. Our approach is based on deep kernel (active) learning, DK(A)L, which combines a deep neural network with a Gaussian process and allows learning relationships between local structure and functionality encoded in spectra on-the-fly. Each measurement informs the subsequent measurement in this setup by increasing the model's knowledge base for the structure-property relationship of interest. It was first demonstrated for the bulk and edge plasmon discovery in STEM-EELS experiments and later applied to rapid studies of symmetry-breaking distortions and internal electric and magnetic fields in 4D STEM experiments on graphene (\autoref{fig:dkl}) and \ce{MnPS3}.
    
    The DKAL experiment starts with featurizing a 2D (HA)ADF image by splitting it into small patches at each pixel coordinate. The patch size can be determined by the characteristic length scale of the phenomena of interest, or it can be chosen ad hoc, as demonstrated in \autoref{fig:dkl}(a). The next step is to determine a scalarizer function, that is, a function that converts a spectral signal (\autoref{fig:dkl}(b)) to the physical property of interest. The scalarizer can, for example, be a peak energy or maximum intensity of selected mode(s) in EELS, or the center of the mass shift in 4D-STEM, as shown in \autoref{fig:dkl}(c). It can also be based on more advanced analyses, such as those involving physics-based inversion of the 4D-STEM data toward the scattering potential with a subsequent selection of the associated features of interest. After featurizing the structural image and defining the scalarizer, we perform several measurements in randomly selected pixel coordinates and use corresponding image patches and scalarized spectra to train a DKL model. The trained model is then used to predict the next measurement point. After performing a measurement at the suggested location, we update our training set and re-run the model training and prediction steps. The process is iterated until the experimental budget is exhausted or the uncertainty in predictions falls to a required level or the required predictability is achieved, as demonstrated in \autoref{fig:dkl}(d).
    
    Based on this scheme, automated experiments in both STEM-EELS and 4D-STEM using DKL were performed in real time on the operational STEM \cite{AE_STEM_EELS, AE_STEM_4D}. NION microscopes were used where the NION Swift control software allows flexible python control of the microscope hardware elements needed to conduct automated experiments\cite{nion_swift1, nion_swift2}. A particular example highlighting the motivation of using automated experiments occurs in \cite{AE_STEM_EELS} where a previously unknown plasmon edge mode in 2D \ce{MnPS3} was discovered using the DKL approach. By scalarizing the EEL signal in such a way as to maximize a ratio of low energy to high energy spectral peaks, the correlation between material edges and this EEL peak ratio was actively learned during the experiment. For beam sensitive specimens like \ce{MnPS3}, automated experiments are even further justified. In the context of 4D-STEM, diffraction patterns were scalarized according to center of mass shifting and correlated to structural features in graphene, where the exploration of the sample was performed by learning where the maximum center of mass shifts are most likely to occur.

	\subsection{\label{ssec:hpc}From the edge computer to high-performance compute clusters}
	\begin{figure*}
		\includegraphics[width=\textwidth]{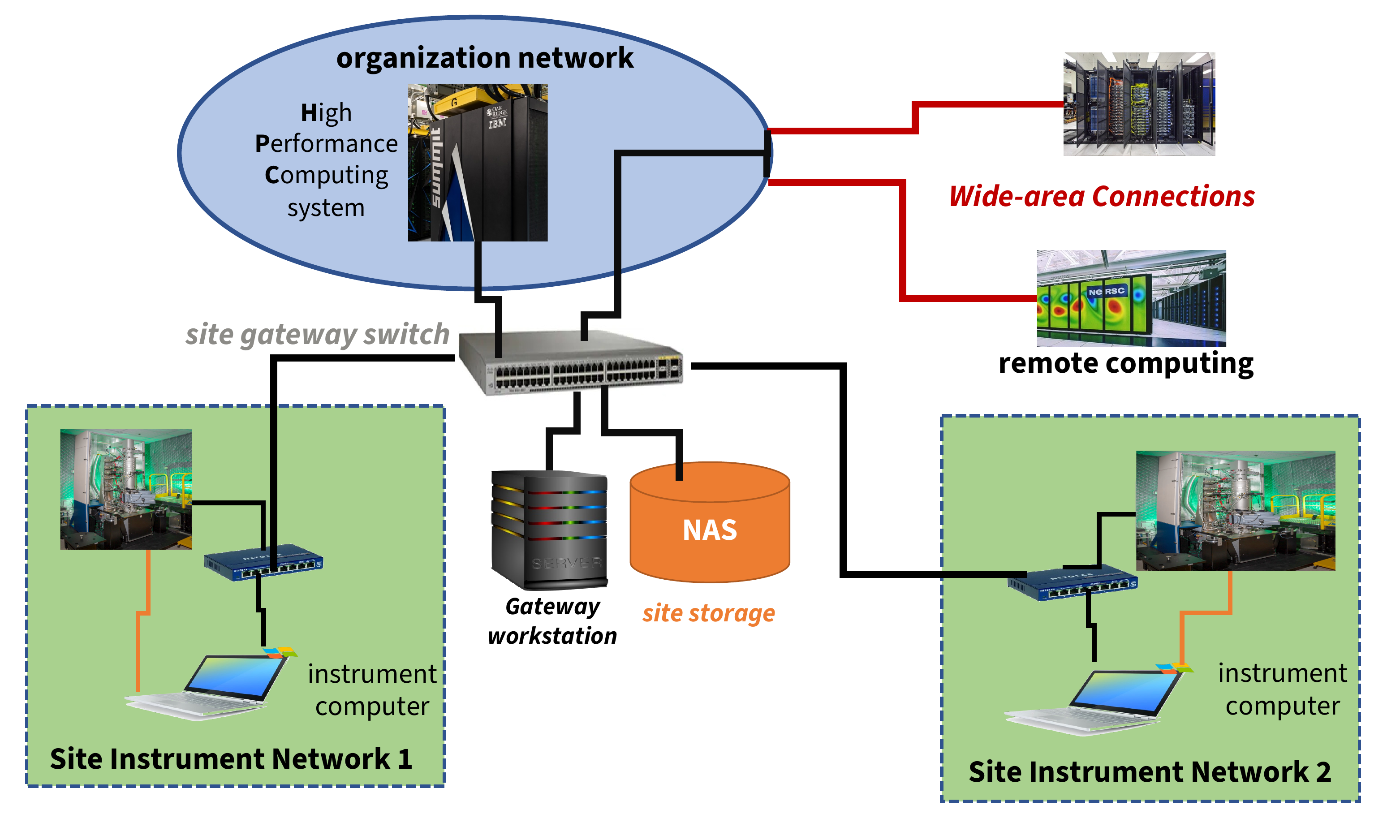}
		\caption{\label{fig:connection}\textbf{Schematic of connection between the microscope and high-performance computing through network assisted storage gateways.} The schematic shown here demonstrates a setup for connecting two separate electron microscopes to a common internal network gateway, along with access to remote HPC systems through wide-area connections.}
	\end{figure*}
	
	Comprehensive studies utilizing an open-access, overall framework capable of directly mapping between experimental observations and computational studies using deep learning approaches are still in their development stages. Here, the primary roadblocks are the difference in time, length scales between two regimes, and associated latencies to model and understand a physical behavior. At ORNL, we have developed a deep learning-based framework that can be useful in addressing such challenges. We focus on employing deep convolutional neural networks to identify atomic features (type and position) in a material. We use features within the AtomAI utility functions to construct a bulk conventional unit cell, supercell, or surface, based on the type of simulations performed and material properties of interest\cite{atomai, ayana_atomai}. 
	
	Post AtomAI initialization, we perform numerical simulations to find the optimized geometry and temperature-dependent dynamics of system evolutions. The outputs, along with associated uncertainties in predictions at various levels, are obtained utilizing this framework and may be subsequently used to evaluate and modify experimental conditions and regions of interest to drive an automated experiment and build a platform that can ``learn'' material properties on the fly during an electron microscopy experiment itself. This entire framework can be performed on-the-fly such that observational data is directly transferred from microscopes to an edge computer, such as the Jetson AGX Xavier. This GPU-based platform is then used to analyze, train, or tune pre-trained DL models, followed by simulations using CPU-based high-performance computing (HPC) resources. Altogether the feedback from measurements, simulations, and DL models are then introduced back to the workflow to guide experiments better while learning from theoretical models\cite{ayana_elit}. 
	
	\begin{figure*}
		\includegraphics[width=\textwidth]{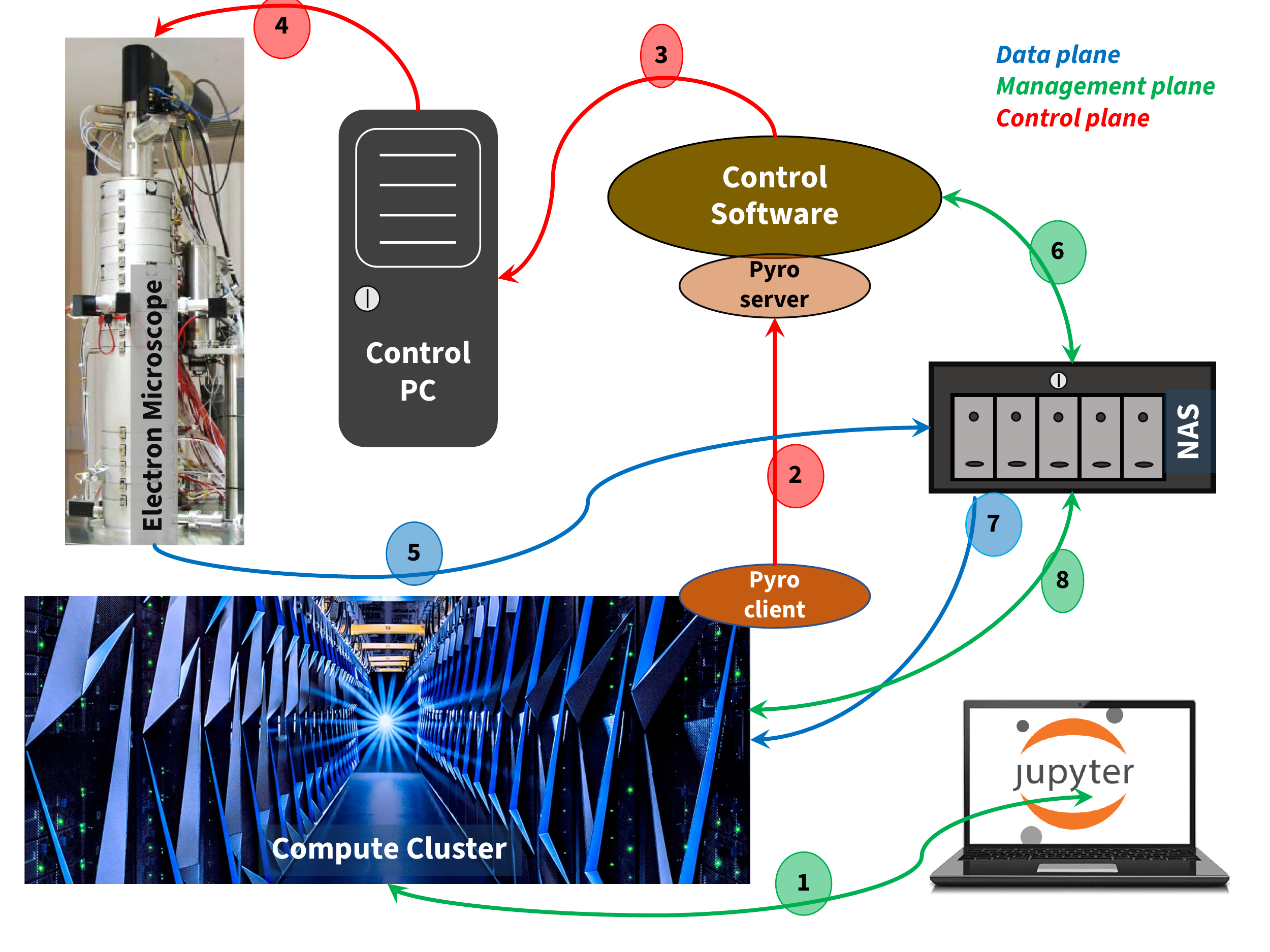}
		\caption{\label{fig:nion_setup}\textbf{Current setup at ORNL for remote access and on-demand microscopy data transfer between remote workstations and NION microscopes.} The communication between the microscopes, control computers, network-assisted storage has eight separate components. The components themselves can be classified into data transfer planes (blue), management plane (green) and control plane (red).}
	\end{figure*}
	\subsection{\label{ssec:hpcdata}Seamless connectivity between microscopes, edge systems and HPC clusters}

	The schematic of the system that is being setup at ORNL is demonstrated in \autoref{fig:connection}. The scheme shown here connects two separate electron microscopes together. Each microscope has a connection between the microscope and the control computer, while the control computer is connected to a gateway workstation and network-assisted storage (NAS) through a site gateway switch. Through this gateway switch, the gateway workstation can be connected to lab-area HPC resources and also remote HPC resources (such as NERSC at Berkeley Lab) through wide-area networks. 

	Recently, our team has been successful in building upon this setup, to deploy a prototype system at ORNL, whose connectivity diagram is demonstrated in \autoref{fig:nion_setup}\cite{ornl_nion_communication, escience}. The workflow is executed as follows: the operator starts a Jupyter session on the data center and enters the desired microscope operation, with the notebook session itself operable from any system on the network with verified credentials, as visualized through process 1. This operation script is transmitted through Pyro client to server running on NION Swift on the microscope control computer, as shown by process 2. NION Swift running on the control computer executes this script through processes 3 and 4. The output data from the microscope detectors is streamed to the NAS (process 5), with the state of the data collection communicated to NION Swift (process 6). Since the NAS is also samba mounted on the cluster, this data is also available immediately, as shown through process 7. As the data is being collected on the NAS, the cluster starts processing the data, with the results being available to the operator as outputs in the Jupyter session. 

	\subsection{\label{ssec:data_compression}Data management through compression and open-source file formats}
	
	One of the challenges in coupling microscopes and compute clusters is that they are often geographically separated; thus, the bottleneck is often network connections that slow down data transfer. Many of these issues can be mitigated by developing physics-informed compression techniques, with the compression routines residing on the edge computer and transferring only compressed data to HPC clusters on the network. Both lossless and lossy compression would be avenues that should be explored. Many current techniques, such as 4D-STEM strain mapping or EELS mapping, which focus on small data sections, would probably benefit tremendously from lossy data compression techniques. 
    
    Additionally, large microscopy datasets are often composed of multiple individual frames, which can be CBED patterns in 4D-STEM or individual frames in cryo-STEM or in-situ TEM. In each of these cases, individual frames are often extraordinarily similar to neighboring frames in a single dataset and are thus low entropy datasets. Arithmetic coding-based lossless compression approaches can be potent in many instances, leading to data compression often by several factors. Compressed data should be stored in open-source data formats, and codes on HPC should be built to work directly on compressed data without decompressing the data on the cluster. 
    
    In the last several years, a few groups have also tackled the issue of data storage formats in the materials community. This has led to some development of open-source microscope data containers capable of performing provenance tracking, such that compressed data and analysis can live together in a single dataset. Work in this area has led to the development of the EMD data format\cite{emdataset}, which is based on the open source HDF5 specification. Another recent work at ORNL has been the Universal Spectroscopic and Imaging Data (USID)\cite{usid, USIDarxiv}, which has been accompanied by the open-source pyUSID package to read and write USID datasets\cite{pyUSID}. 
    
    The future microscopy system should thus directly stream data from detectors to GPUs on the connected edge system, where online compression can be performed on the data being streamed in. Alongside this, compressed data should be streamed to HPC systems, where the data is processed and stored in open-source data containers, and the processed results, which are often significantly more sparse, streamed back to the microscope operator. The processed data can also be fed into decision-making algorithms, which can be used to drive the microscope for results-guided automated experimentation. This setup will also, however, lead to networking challenges, as data, decisions, and output will be streamed simultaneously and will probably need the development of reliable switching and networking protocols to couple microscopy and computation together.

	\subsection{\label{ssec:vtwin}Continuous provenance tracking through digital twins on connected edge and HPC systems}
	As mentioned in \autoref{ssec:hpcdata}, one of the goals for the development the data containers is storage of metadata in machine-readable formats and provenance tracking. While the microscope data files itself are often multidimensional arrays, making sense of them requires access to metadata. The authors' experience in this often shows that current vendor-generated metadata is often incorrect, and needs significant manual intervention to correct and clean such data. Secondly, over the course of an experimental session the microscope parameters often ``drift'', and thus the metadata that may have been stored reflect the initial starting conditions, while later experiments may be significantly different.

	Yet, a modern microscope often has hundreds of individual sensors that generate information about individual lens and holder parameters, often in real time. One way to make sense of this data would be to build virtual infrastructure twins, where a simulation of the microscope is running on a cluster, which is updated in real time based on individual lens values. Such a system, can not only then save the digital twin of a microscope in real time providing detailed provenance tracking at every time point of an experiment, but can also point out errors in metadata if the observed sensor outputs don't match up. Notably, a setup like this would completely eliminate the need for a skilled operator to verify the accuracy of a metadata, and generate the state of the microscope while analyzing the data, allowing for significantly more complex autonomous experiments where the microscope's parameters are changing during the course of an experiment.

	Additionally, continuous state monitoring opens up the door for Bayesian predictive experiments. Whether it is for inverse problems such as ptychography, or for in-situ experiments, provenance tracking allows for completely automated, predictive science. This data is already available today, but it's rarely collected in a single location in a machine-parsable format. In-situ holders for heating, biasing or chemical reactions track the states continuously, while most modern microscopes track individual lens currents and settings. However, as far as the authors know these datasets have not been used till data to build a digital microscope twin. Building a highly accurate digital twin also removes the need for multiple experiments. A future system can then run a complete virtual experiment, and design a real-world microscope to find the data needed.

	\section{\label{sec:conclusion}Conclusions}

	As the challenges and opportunities we demonstrated in the roadmap above demonstrate, this is currently an extraordinarily exciting time for electron microscopy. Two fields, data science and microscopy are colliding together, and the future microscope will be not be a standalone system, but a multi-purpose scientific tool tightly coupled to local edge and remote HPC computational facilities. The current divisions in physical sciences, between materials synthesis, characterization and theory -- are not the future. All three aspects, can be and should be present together in a single setup, that can ``learn the physics'' directly from the experimentation itself -- autonomously -- literally, a ``Lab on a Beam''\cite{lab_beam}.

	\section*{Acknowledgements}

	This research is sponsored by the INTERSECT Initiative as part of the Laboratory Directed Research and Development Program of Oak Ridge National Laboratory, managed by UT-Battelle, LLC, for the US Department of Energy under contract DE-AC05-00OR22725.  This research used resources of the Center for Nanophase Materials Sciences, which is a DOE Office of Science User Facility.  
	
	\bibliography{A_Roadmap}
\end{document}